\documentclass[12pt]{article}
\usepackage{graphicx}

\newcommand\pubnumber{SNSN-323-63}
\newcommand\pubdate{\today}

\def\institute{Laboratoire de Physique Nucl\'eaire et de Hautes Energies (LPNHE), UPMC, Universit\'e Paris-Diderot, CNRS/IN2P3, Paris, France}
\def\support{\footnote{On behalf of the ATLAS Collaboration}}

\def\Title#1{\begin{center} {\Large #1 } \end{center}}
\def\Author#1{\begin{center}{ \sc #1} \end{center}}
\def\Address#1{\begin{center}{ \it #1} \end{center}}

\newcommand\pubblock{\rightline{\begin{tabular}{l} \pubnumber\\
         \pubdate  \end{tabular}}}
\newenvironment{Abstract}{\begin{quotation}  }{\end{quotation}}
\newenvironment{Presented}{\begin{quotation} \begin{center} 
             PRESENTED AT\end{center}\bigskip 
      \begin{center}\begin{large}}{\end{large}\end{center} \end{quotation}}

\def\beq{\begin{equation}}
\def\eeq#1{\label{#1}\end{equation}}
\def\eeqn{\end{equation}}

\def\beqa{\begin{eqnarray}}
\def\eeqa#1{\label{#1}\end{eqnarray}}
\def\eeqan{\end{eqnarray}}

\let\bar=\overbar

\def\Dslash{\not{\hbox{\kern-4pt $D$}}}
\def\dslash{\not{\hbox{\kern-2pt $\del$}}}

\def\msb{{\bar{\ssstyle M \kern -1pt S}}}

\begin{document}
\begin{titlepage}
\pubblock

\vfill 
\Title{Latest ATLAS measurements of top quark properties using ${\rm t{\bar t}}$ events.}
\vfill
\Author{Fr\'ed\'eric Derue\support}
\Address{\institute}
\vfill
\begin{Abstract}
Recent measurements of top quark properties with ${\rm t{\bar t}}$ events
collected by the ATLAS experiment in proton-proton collisions at the center
of mass energy $\sqrt{s}=8$ and $13$~TeV, are presented.
The measurements of the top quark width, spin and spin correlations,
polarisation, the \ensuremath{W} boson helicity fractions, charge and CP asymmetries
are discussed.
Finally, recent results obtained on colour flow effects in ${\rm t{\bar t}}$
events are presented.
All the measurements are consistent with the Standard Model expectations.
\end{Abstract}
\vfill
\begin{Presented}
$10^{th}$ International Workshop on Top Quark Physics\\
Braga, Portugal,  September 17--22, 2017
\end{Presented}
\vfill
\end{titlepage}
\def\thefootnote{\fnsymbol{footnote}}
\setcounter{footnote}{0}
\section{Introduction}
The top quark is the heaviest fundamental particle in the Standard Model (SM)
of particle physics.
It has a lifetime shorter than the hadronisation timescale and gives the
unique opportunity to study the properties of a “bare” quark.
The huge top quark sample collected by the ATLAS
experiment~\cite{ATLASdetector} at the Large Hadron Collider~\cite{LHC} at CERN
allow to perform precision tests of the SM.
The measurements reported here use events collected by the ATLAS
detector in proton-proton collisions at the center
of mass energy $\sqrt{s}=8$ and $13$~TeV in which a ${\rm t{\bar t}}$ pair
is produced.
Monte Carlo (MC) simulated event samples with full detector simulation
are used to model the signal and backgrounds~\cite{simul}.
\section{Top quark polarisation and ${\rm t{\bar t}}$ spin correlation}
Top quark pairs produced by the strong interaction have essentially
unpolarised spins.
However, the spins of the top quark and top anti-quark are expected to be
correlated.
As the top quark decays before it hadronises, information about the top quark
spin is passed on to its decay products and the strength of the
${\rm t{\bar t}}$ spin correlation can be measured using angular distributions.
The measurement of 15 spin observables is performed using
$20.2~{\rm fb^{-1}}$ of $\sqrt{s}=8$~TeV collision data in the dilepton
channel~\cite{spindilepton}.
The normalised double-differential cross section for ${\rm t{\bar t}}$
production and decay is:
\begin{equation}
  \frac{1}{\sigma}\frac{d^2\!\sigma}{d\!\cos\theta_{+}^{a} d\!\cos\theta_{-}^{b}} =
  \frac{1}{4}\left( 1 + B_{+}^{a} \cos\theta_{+}^{a} + B_{-}^{b} \cos\theta_{-}^{b}
  - C(a,b) \cos\theta_{+}^{a} \cos\theta_{-}^{b} \right),
\end{equation}
where $B^{a,b}$ and $C^{a,b}$ are the polarisation and spin correlation
coefficients along spin quantisation axes $a$ and $b$ respectively and
$\theta_{+}^{a}$ ($\theta_{-}^{b}$) is the angle between the positive (negative)
lepton and quantisation axis $a$ ($b$).
Different spin quantisation axes can be chosen, each with a different
expected spin correlation or polarisation strength.
The most commonly chosen axes used is the helicity axis, $k$, in which the
spin quantisation axis is chosen as the direction of the top quark in the
${\rm t{\bar t}}$ system rest frame.
The angle $\theta$ is then formed using this axis and the direction of the
charged lepton in the top quark rest frame.
$B^a = 3<\cos\theta^a>$ are 6 polarisation coefficients,
and $C(a,b) = −9<\cos\theta_{+}^{a} \cos\theta_{-}^{b}>$ are 9 spin correlation
coefficients.
All measured coefficients are in good agreement with the SM theoretical
expectations, within uncertainties.
The largest contribution to the uncertainty may vary for the different
measured parameters, in general it is due to ${\rm t{\bar t}}$ signal modeling
and accounts for approximately 80\% of the total systematic uncertainty.
\section{Charge asymmetry}
Production of ${\rm t{\bar t}}$ pairs is predicted to be symmetric under charge
conjugation at leading order in quantum chromodynamics (QCD) in the SM.
However, at next-to-leading order a small ($\simeq$1\%) charge asymmetry is
introduced resulting in the rapidity distribution being slightly broader for
top quarks than for top anti-quarks.
This asymmetry is passed on to the decay products of the top and top
anti-quark and thus also exists between the charged leptons in dilepton
${\rm t{\bar t}}$ events.
Although the ${\rm t{\bar t}}$ charge asymmetry is predicted to be small
in the SM, beyond the SM physics can cause it to be enhanced.
The ${\rm t{\bar t}}$ charge asymmetry is defined as:
\begin{equation}
A_{C}^{{\rm t{\bar t}}} = \frac{N(\Delta|y|>0) - N(\Delta|y|<0)}{N(\Delta|y|>0) + N(\Delta|y|<0)},
\end{equation}
where $\Delta|y|=|y_t|-|y_{{\bar t}}|$, while the lepton-based charge asymmetry
is defined as:
\begin{equation}
A_{C}^{l^+l^-} = \frac{N(\Delta|\eta|>0) - N(\Delta|\eta|<0)}{N(\Delta|\eta|>0) + N(\Delta|\eta|<0)},
\end{equation}
where $\Delta|\eta|=|\eta_{l^+}|-|\eta_{l^-}|$.
The measurement of these asymmetries is performed
using $20.3~{\rm fb^{-1}}$ of $\sqrt{s}=8$~TeV collision data in the
lepton+jets and dilepton channels~\cite{chargeasymmetry1,chargeasymmetry2}.
In addition to the inclusive measurements of $\Delta|y|$ and $\Delta|\eta|$,
these asymmetries are measured versus
other variables, such as the invariant mass of the ${\rm t{\bar t}}$
system, $m_{{\rm t{\bar t}}}$.
Beyond the SM physics may enhance the ${\rm t{\bar t}}$ charge asymmetry at
high $m_{{\rm t{\bar t}}}$ for example due to the presence of a heavy
resonance decaying to a ${\rm t{\bar t}}$ pair.
In all cases, the data are well modelled by the SM prediction and consistent
with only a small charge asymmetry.
\section{$CP$ asymmetries}
$CP$ violation can be probed using weakly decaying $B$-hadrons from top quark
decays using $20.3~{\rm fb^{-1}}$ of $\sqrt{s}=8$~TeV collision
data in the lepton+jets channel~\cite{topCP}.
The charge of the lepton is used to determine the charge of the $b$-quark at
production, $\alpha$.
Events are then required to contain an additional muon associated to a
$b$-tagged jet, this is known as a “soft muon” and is used to determine the
charge of the $b$-quark when it decays, $\beta$.
Given the number of events with charges $\alpha$ and $\beta$, $N^{\alpha\beta}$, 
same-sign, $A_{SS}$, and opposite-sign, $A_{OS}$, charge asymmetries are
constructed:
\begin{eqnarray}
A_{SS} &=& \frac{(N^{++}/N^+) - (N^{--}/N^-)}{(N^{++}/N^+) + (N^{--}/N^-)} \, ,
A_{OS} = \frac{(N^{+-}/N^+) - (N^{-+}/N^-)}{(N^{+-}/N^+) + (N^{-+}/N^-)},
\end{eqnarray}
where $N^{+(-)}$ is the number of events in which the charged lepton has
positive (negative) charge.
The measured values are 
%
$A_{SS} = −0.007 \pm 0.006\, ({\rm stat.}) \pm 0.002\, ({\rm expt.}) \pm 0.005\, ({\rm model})$ and
$A_{OS} = 0.0041 \pm 0.0035\, ({\rm stat.})\,^{+0.0013}_{-0.0011}\, ({\rm expt.}) \pm 0.0027\, ({\rm model})$,
%
consistent with zero.
From these measurements, four CP asymmetries (one mixing and three direct) are
measured and are found to be compatible with zero and consistent with the
SM expecations.
\section{$W$ helicity measurement in top quark events}
The $W$ boson from the top quark decay has only left-handed and longitudinal
polarisation, the right-handed polarisation is heavily suppressed.
The corresponding helicity fractions are determined by the $Wtb$ vertex
structure and are measured using $20.2~{\rm fb^{-1}}$ of $\sqrt{s}=8$~TeV
collision data in the lepton+jets channel~\cite{Whelicity}.
They are extracted from the angular distributions of the
$W$ decay products through the normalised differential cross section:
\begin{equation}
  \frac{1}{\sigma}\frac{d\!\sigma}{d\!\cos\theta^{*}} =
  \frac{3}{4}\left( 1 -\cos^2\theta^{*} \right) F_0 +
  \frac{3}{8}\left( 1 -\cos\theta^{*} \right)^2 F_L +
  \frac{3}{8}\left( 1 +\cos\theta^{*} \right)^2 F_R ,
\end{equation}
where $\theta^*$ is the angle between the direction of flight of the top quark
in the $W$ rest frame and the direction of flight of the lepton or the
$d$-type quark for the leptonic or hadronic $W$ decay, respectively.
Templates correspondings to the $\cos\theta^{*}$ distributions for the three
helicity components, and for both hadronic and leptonic top decays, are used
to obtain the best description of the measured distributions, and to extract
the helicity fractions.
The measurement on the leptonic side is the most sensitive since the hadronic
side is affected by a worse separation power and by larger uncertainties.
The fitted fractions of longitudinal, left- and right-handed polarisation
states are $F_0=0.709\pm 0.019$, $F_L=0.299\pm 0.015$ and
$F_R=-0.008\pm 0.014$.
No significant deviation from SM prediction is observed, thus limits on
anomalous couplings of the $Wtb$ vertex are also set.
\section{Direct measurement of the top quark decay width}
\begin{figure}[h!]
\centering
\includegraphics[width=0.4\textwidth]{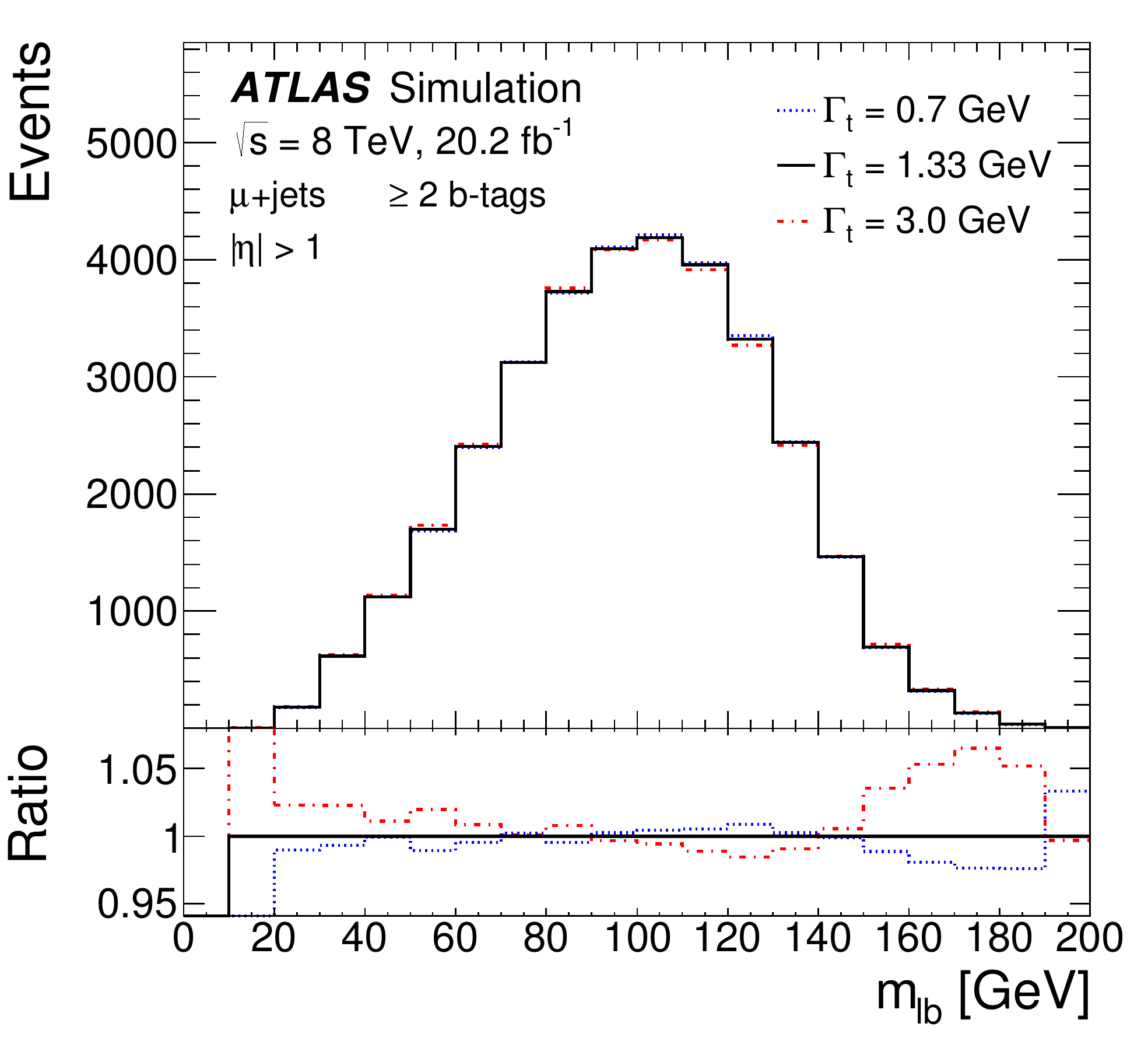}
\caption{Templates for $m_{lb}$, the reconstructed invariant mass of the
$b$-jet of the semileptonically decaying top quark and the corresponding
lepton~\cite{topwidth}.}
\label{fig:topwidth}
\end{figure}

The first direct measurement of the top quark decay width is done using
$20.2~{\rm fb^{-1}}$ of $\sqrt{s}=8$~TeV
collision data in the lepton+jets channel~\cite{topwidth}.
Two variables sensitive to decay width of top quark are used:
$m_{lb}$ - invariant mass of lepton and corresponding $b$-jet from top quark
decay (see Fig.~\ref{fig:topwidth})
and $\Delta R_{{\rm min}}(j_{\rm b},j_{\rm light})$ - $\Delta R$
between $b$-jet from hadronically decaying top and
closest light jet from hadronically decaying $W$ boson.
Events are split by lepton flavour (e or $\mu$), number of $b$-tagged jets
(exactly one $b$-tagged jet and at least two b-tagged jets) and jet $|\eta|=1$
to decrease systematic uncertainties in jet energy reconstruction.
Templates with different widths are fitted to data and the measured width is:
\begin{equation}
\Gamma_{\rm{top}}= 1.76 \pm 0.33\, ({\rm stat.})\,^{+0.79}_{-0.68}\, ({\rm syst.})~{\rm GeV} \,
\end{equation}
assuming top mass $m_t=172.5$~GeV.
Measurement is limited by jet energy reconstruction and signal modelling
uncertainties.
The measured width is in good agreement with the SM prediction.
\section{Colour flow}
\begin{figure}[h!]
\centering
\includegraphics[width=0.4\textwidth]{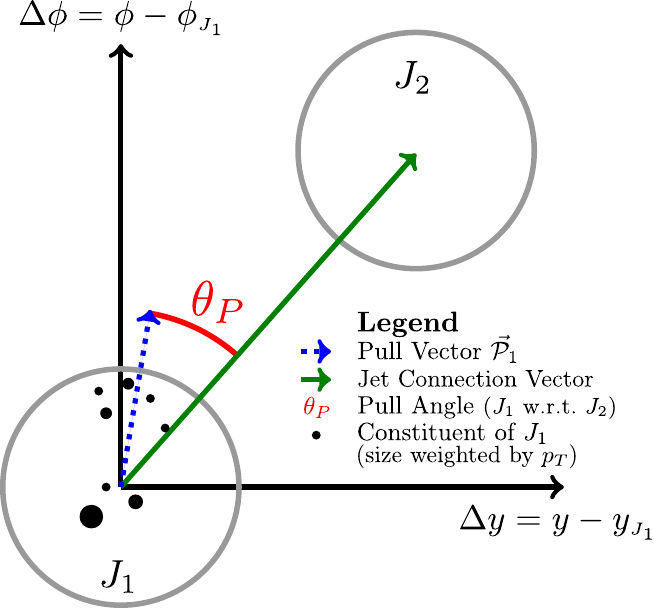}
\caption{Jet pull observables for a dijet system. The pull angle $\theta_P$
is between the pull vector and the vector connecting
two jets~\cite{colourflow}.}
\label{fig:colourflow}
\end{figure}

In nature only colour neutral hadrons can be observed, and direct measument
of the quantum chromodynamics (QCD) interaction of their constituents
is not possible.
Quark and gluons are measured as jets, but colour connection influences jet
form, structure and event topology.
An observable expected to carry colour information about a jet is the pull
vector, a $p_{\rm T}$-weighted radial moment of the jet,
computed from its constituents.
For a dijet system is the pull angle, between the pull vector and the vector
connecting the two jets (see Fig.~\ref{fig:colourflow}).
In the lepton+jets event topology, the jets originating from the hadronic
$W$ decay are colour connected, while the two $b$-tagged jets are not.
These observables are studied using $36~{\rm fb^{-1}}$ of $\sqrt{s}=13$~TeV
collision data in the lepton+jets channel~\cite{colourflow}.
After bin-by-bin background subtraction, and removing effects of the detector
smearing as obtained from simulation, the signal distributions are then
unfolded to a particle-level spectrum, using an Iterative Bayesian method.
As a result, a general good agreement can be found with SM predictions,
even if observables sensitive to colour flow remain poorly modelled.
\section{Conclusion}
The large number of ${\rm t{\bar t}}$ pairs collected by the ATLAS experiment
allows to make precise measurements of top quark properties.
All top quark properties, measured using the
data collected in LHC Run~1 and the first round of Run~2,
are in good agreement with the SM expectations.


\end{document}